\pdfoutput=1
\documentclass[singlespacing]{elsart}
\usepackage{graphics}
\usepackage{graphicx}
\usepackage{epsfig}
\usepackage{amssymb}
\begin{document}
\begin{frontmatter}
\title{Construction of Solenoidal Immersed Velocity Vectors Using the Kinematic Velocity--Vorticity Relation}
\author{Fereidoun Sabetghadam \corauthref{cor1}}
\ead{fsabet@srbiau.ac.ir}
\author{Shervin Sharafatmandjoor, Mehdi Badri}
\author{}
\corauth[cor1]{}
\address{Mechanical and Aerospace Eng. Dept., Science and Research Branch, Islamic Azad University (IAU), Tehran, Iran}
\begin{abstract}
The present paper suggests a method for obtaining incompressible solenoidal velocity vectors that satisfy approximately the desired immersed velocity boundary conditions. The method employs merely the mutual kinematic relations between the velocity and vorticity fields (i.e, the curl and Laplacian operators). An initial non-solenoidal velocity field is extended to a regular domain via a zero-velocity margin, where an extended vorticity is found. Re-calculation of the velocities (subjected to appropriate boundary conditions), yields the desired solenoidal velocity vector. The method is applied to the two- and three-dimensional problems for the homogeneous Dirichlet, as well as periodic boundary conditions. The results show that the solenoidality is satisfied up to the machine accuracy for the periodic boundary conditions (employing the Fourier--spectral solution method), while an improvement in the solenoidality is achievable for the homogenous boundary conditions.
\end{abstract}

\end{frontmatter}
\section{Introduction}
The Navier-Stokes Equations (NSE) are derived for a continuum. Therefore, many difficulties arise in dealing with any kind of discontinuity,  including finite-size solution domains (simply-connected or multiply-connected). Particularly in an incompressible flow, the problem usually appears as non-solenoidal velocity fields. Consequently, satisfying the mass conservation has been one of the most challenging issues of almost all algorithms of incompressible flow simulations, whether in the primitive variables formulations or the other formalisms. As it is revealed,  the boundary conditions have a vital role here; and therefore, appropriate definition of the boundary conditions received significant attention in the literature \cite{Gresho, Quarta93, Rempfer}.\\
The literature of the subject is too rich to completely be summarized in this paper (see e.g. \cite{Rempfer} and the references in there);  however, as the most pertinent works, we shall make reference to the works of Quartapelle and Valz-Gris \cite{Quarta81}, and also Quartapelle \cite{Quarta93}, which studied the issue extensively. Calhun \cite{Calhoun} used the immersed boundary technique in the vorticity--stream function formulation, and introduced an appropriate distribution of the vorticity in order to satisfy the overall mass balance. From another viewpoint, the problem was studied in relation to the compatibility conditions (also called  the space--time corner singularity). The required compatibility conditions for the NSE was studied by Temam \cite{Temam1}, Gresho and Sani \cite{Gresho}, and also by Boyd and Flyer \cite{Boyd}, and then by Flyer and Swarztrauber \cite{Flyer}).\\
The main contribution of the present paper is to suggest a method for construction of a {\it nearly} solenoidal velocity field, which satisfies {\it approximately} the desired {\it immersed} velocity boundary conditions, using the kinematic velocity--vorticity relation. The central idea is to use the well-known property that the divergence of the velocities, obtained from the velocity--vorticity equation is a harmonic function; and therefore, gets its maximum values on the boundaries. Consequently, the divergence  can be reduced (or even vanished) all over the field, by restricting (or vanishing) the divergence near the boundaries.  In the present method, these restrictions are implemented by considering a zero-velocity margin in the vicinity of the outer regular boundaries.\\
Embedding non-regular solution domains in a bigger regular domain was shown to be useful in solution of the scalar Poisson's problems \cite{Boyd2,Sabetghadam1}. Now the present work may be seen as an extension of the method to the vector Poisson's problems that the solenoidal solutions are desired. As it will be shown, a solenoidal velocity field (within the machine accuracy) is achievable for the periodic boundary conditions (that is, the no-boundary problems), and also improvements in the solenoidality is possible for the homogeneous Dirichlet boundary conditions. The method can be used in recently popularized one-fluid \cite{Tryggvason}, or one-continuum \cite{Sugiyama} immersed boundary methods, as it was used successfully our previous work \cite{Sabetghadam2}. \\
The paper is continued by description of the suggested algorithm, follows by a section contains our arguments about the reasons that the method works (that is, $\S$ \ref{sec3}). As our numerical experiments, two- and three-dimensional fixed and moving boundary problems are examined, with homogeneous as well as periodic boundary conditions.
\newpage
\section{The suggested procedure}
Given an arbitrary (may be non-solenoidal) velocity vector ${\rm{\bf u}}^*=(u^*_1,\cdots, u^*_d) \in {\cal L}^2( \Omega_f)$, defined on the flow domain $\Omega_f\in {{\Bbb R}^d}$, with arbitrary boundary $\Gamma_{\cal B}$, where $d=2,3$; one can  find the corresponding vorticity vector ${\omega}^*=\nabla\times{\rm{\bf u}}^*$ on the $\Omega_f$. Now, it is well-known that solution of the Poisson equations
\begin{equation}\label{Eq:1}
\nabla^{2}{\bf u}=-\nabla\times\omega^*,
\end{equation}
on $\Omega_f$ implies
\begin{equation}\label{Eq:3}
\nabla^{2}(\nabla\cdot {\bf u})=0;
\end{equation}
which means  although $\nabla\cdot {\bf u}$ is not necessarily zero, it is a harmonic function, and therefore, gets its maximum values on the boundaries. The above statement has been demonstrated several times by many authors \cite{Quarta81,Quarta93}, and motivated many attempts in obtaining solenoidal velocities by vanishing the divergence right at the boundaries \cite{Calhoun, E}.\\
\begin{figure}[t]
\setlength{\unitlength}{1mm}
\centerline{\includegraphics[width=7.5cm]{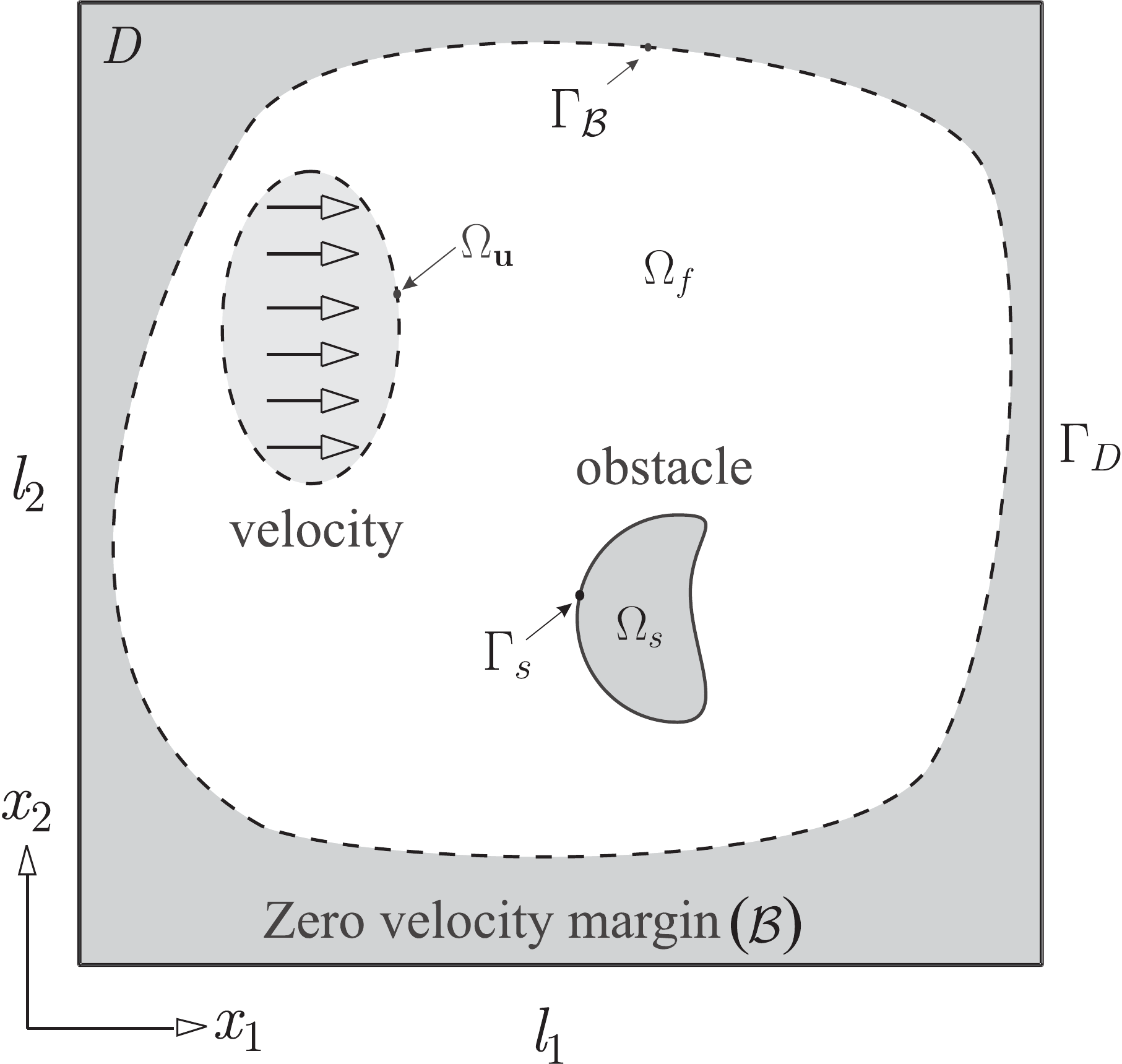}}
\caption{The flow domain $\Omega_f$, together with fixed or moving obstacle(s) $\Omega_s$, and other given-velocity regions $\Omega_{\rm{\bf u}}$, are embedded in the regular solution domain $D$ (with regular boundary  $\Gamma_{D}$), via a zero-velocity margin $\cal B$.}
\label{fig1}
\vspace{7mm}
\end{figure}
Now, we suggest limiting the divergence inside the solution domain via controlling the divergence in the vicinity of the outer boundaries (not merely at the boundaries), via definition of a zero-velocity margin. According to Fig. \ref{fig1}, The regular solution domain $\bar{D}=D+\Gamma_D$ contains all the fluid--solid system, including the moving regions of the fluid with given velocity boundary
conditions (that is $\Omega_{\bf{\rm u}}$), in addition to the fixed or moving immersed solid objects. Now this fluid--solid system is embedded in a regular domain $D$ via a zero-velocity margin $\cal B$.\\
With the above definitions in mind, the following procedure is suggested:
\begin{figure}[t]
\centering
\includegraphics[width=0.4\textwidth]{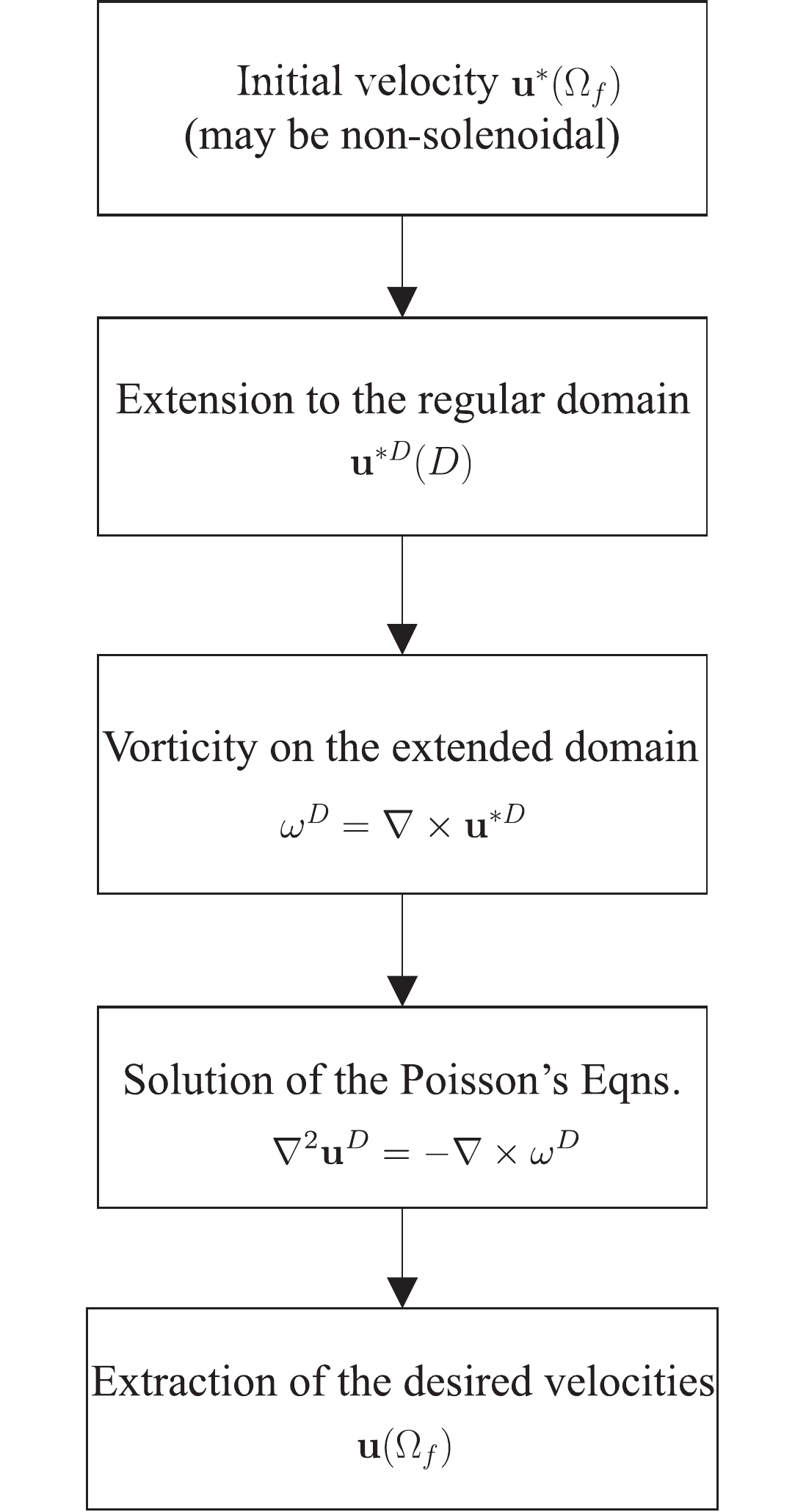}
\caption{Flowchart of the proposed algorithm. The initial velocities on the flow domain $\Omega_f$ is extended to a bigger regular domain $D$, where the extended vorticity $\omega^D$ is obtained and the Poisson's equations are solved on. Then the modified velocities $\rm{\bf u}$ are extracted from $\rm{\bf u}^D$. }
\label{fig2}
\vspace{5mm}
\end{figure}
\begin{itemize}
\item[{\bf i)}] Given the arbitrary (may be non-solenoidal) velocity vector ${\rm{\bf u}}^*(\Omega_f)$ defined on $\Omega_f$, the extended velocity ${{\rm{\bf{u}}}^{*D}}(D)$, defined on the regular domain $D$, is found via addition of the zero-velocity margin $\cal B$, and other immersed regions ($\Omega_{\rm{\bf u}}$ and others) to the ${\rm{ \bf u}}^*$.
\item[{\bf ii)}] The vorticity $\omega^D=\nabla\times{\rm{\bf u}}^{*D}$ is obtained on $D$.
\item[{\bf iii)}] The modified extended velocity vector ${\rm{\bf u}}^D$  is obtained from solution of the Poisson's equations
\begin{equation}
\nabla^{2}{\bf u}^D=-\nabla\times\omega^D \quad \quad {\bf x}\in D \label{Poiss2}
\label{Poissons}
\end{equation}
subjected to some boundary conditions on $\Gamma_D$, which will be discussed in the next section.
\item[{\bf iv)}] The desired modified velocity vector ${\rm{\bf u}}({\Omega}_f)$ is obtained by discarding ${\rm{\bf u}}^D(\cal B)$ and the other immersed regions ($\Omega_{\rm{\bf u}}$ and others) .
\end{itemize}
The above procedure is illustrated in Fig. \ref{fig2}.
\begin{rem}
In fact for the multiply connected domains, the extremum of the divergence may occur on the boundaries of the holes (e.g. ${\Omega}_{\rm{\bf u}}$ or others). However, in the present work we followed an immersed boundary approach, that is,  ${\rm{\bf u}}^{*D}$ includes the velocities of all the holes of $D$. In this way, the extremum of the divergence occurs on $\Gamma_D$, and therefore, it is just needed to control the divergence on $\Gamma_D$.
\end{rem}
\section{On the Solenoidality of ${\rm{\bf u}}$} \label{sec3}
We shall examine the solenoidality of ${\rm{\bf u}}$. To this end, we consider two different boundary conditions for Eq. (\ref{Poissons}), that is, the periodic boundary conditions, which is suitable for the Fourier spectral solvers, and the {\it homogenous} Dirichlet boundary conditions, which can be implemented easily in the finite-difference or finite-volume solvers.
\subsection{Periodic Boundary Conditions}
In the case of periodic boundary conditions (that is, no-boundary problems), equation (\ref{Poissons}) can be transformed into the Fourier space, which leads to
\begin{equation}
\hat{{\rm{\bf u}}}^D=-i\frac{{\rm{\bf k}}}{|{\rm{\bf k}}|^2}\times\hat{\omega}^D,
\end{equation}
in which $\hat{(\cdot)}={{\rm{\bf FT}}}(\cdot)$; and $\rm{\bf k}=(k_1,\cdots,k_d)$ stands for the wavenumber vector, and $i=\sqrt{-1}$. Obviously, $\hat{{\rm{\bf u}}}^D$ is perpendicular to the wavenumbers vector ${\rm {\bf k}}$, that is
\begin{equation}
{\rm{\bf k}}\cdot\hat{{\rm{\bf u}}}^D=0,
\end{equation}
which is a translation of solenoidality of ${\rm{\bf u}}^D$ in the Fourier space \cite{Canuto}. Now the desired (solenoidal) velocity vector ${\rm{\bf u}}$ can be obtained simply by ignoring the solution in the $\cal B$ and the other immersed regions. More than the above arguments, our numerical experiments have also confirmed that in this case ${\rm{\bf u}}$ is solenoidal within the machine accuracy. In fact, achieving perfect conservation of mass for the no-boundary problems is not so surprising, because many assumptions that the NSE are based on are satisfied \cite{Foias}.
\subsection{Homogenous Dirichlet boundary conditions}
For the homogenous boundary conditions on ${\Gamma}_D$, the modified extended velocity ${\rm{\bf u}}^D$ is not strictly solenoidal. However, since  $\nabla^2(\nabla\cdot{\rm{\bf u}}^D)=0$, the divergence of ${\rm{\bf u}}^D$ gets its extremum on the boundary ${\Gamma}_D$. In the other words, the extremum of divergence occurs in $\cal B$, and therefore, discarding the solution in $\cal B$ reduces the divergence such that
\begin{equation}
||\nabla\cdot{\rm{\bf u}}||_{\infty}\leq||\nabla\cdot{\rm{\bf u}}^D||_{\infty}.
\end{equation}
In fact, the maximum values of $\nabla\cdot{\rm{\bf u}}^D$ are discarded  by cutting the solution in $\cal B$; and therefore, the divergence of ${\rm{\bf u}}$ is less than the divergence of ${\rm{\bf u}}^D$.\\
As it will be seen in our numerical experiments, finite-difference solution of Eqns. (\ref{Poissons}) with homogeneous boundary conditions, modifies the solenoidality of an initially non-solenoidal velocity field.
\newpage
\section{Numerical Experiments}
In this section, the proposed algorithm is assessed via examination of some fixed, as well as moving boundary problems in two and three dimensions. We implemented the method both in the Fourier-spectral, and the finite-difference solutions.
\subsection{The modified Chandrashkar--Reid flow}
In the study of the orthogonal basis functions that satisfy the no-slip conditions, Chandrashkar and Reid proposed a velocity field that  satisfies the no-slip boundary condition in one direction, and is periodic in the other direction \cite{Chand}. Then this flow was appeared in the work of Rempfer \cite{Rempfer}. In the light of their works, it is convenient here to construct a velocity field,  which satisfies the no-slip conditions in both directions. Tho this end, we first introduce a one-dimensional characteristic function
\begin{figure}[t]
\centering
\includegraphics[width=0.9\textwidth]{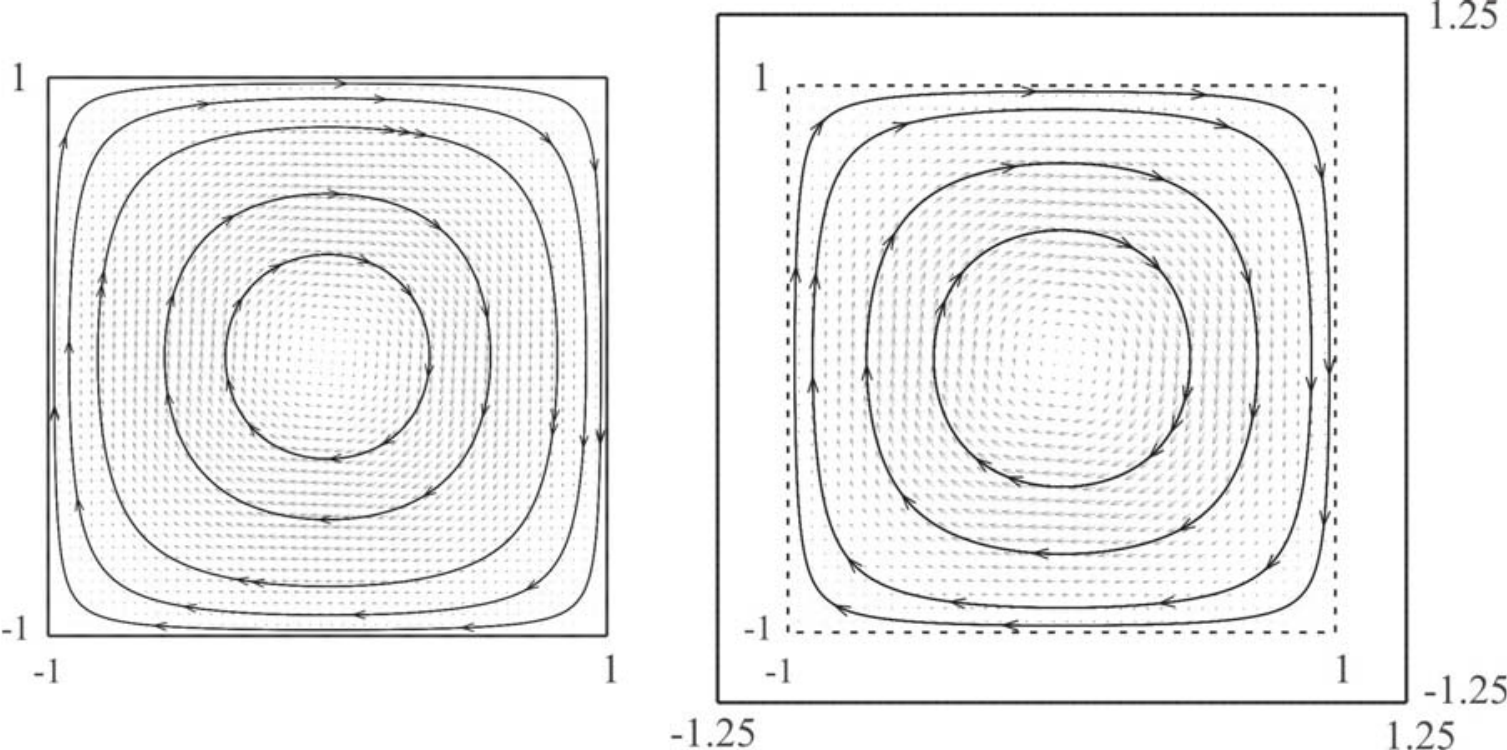}
\caption{{\bf Left:} The modified Chandrashkar--Reid flow, defined on $\bar{\Omega}_f=||{\bf x}||_{\infty}\leq 1$. The no-slip condition is satisfied formally on the boundaries $\Gamma_{\cal B}=\{{\bf x} , ||{\bf x}||_{\infty}=1\}$. {\bf Right:} The solution domain is extended to $\bar{D}=\{ {\bf x}, ||{\bf x}||_{\infty}\leq 1.25 \}$, via addition of the zero velocity margin ${\cal B}=\bar{D} \setminus \bar{\Omega}$.}
\label{vels_fig0}
\vspace{7mm}
\end{figure}
\begin{equation}\label{Eq:6}
\phi{(x_i)}=\cos(\lambda x_i)+A_{\lambda}\cosh (\frac{\pi}{2} x_i); \quad \quad  x_i\in [-1,+1],
\end{equation}
where  $\lambda=2.64$, and $A_\lambda=-\cos\lambda/\cosh (\pi/2)$ are chosen such that $\phi(\pm1)=0$, within the machine accuracy. Using this characteristic function, we construct the desired streamfunction
\begin{equation}\label{Eq:7}
\psi(x_1,x_2)=\frac{1}{G} \phi(x_1) \phi(x_2)
\end{equation}
in which the normalization factor $G$ is chosen such that the unit maximum velocity occurs in the field, that is
\begin{equation}
G=(1+A_{\lambda})[\frac{\pi}{2	}A_{\lambda}\sinh(\frac{\pi}{4})-\lambda \sin(\frac{\lambda}{2})].
\end{equation}
Now the velocities are directly obtainable from this streamfunction as
\begin{eqnarray}\label{Eq:8}
u_1&=&\frac{\partial}{\partial x_2}\psi(x_1,x_2)=\frac{1}{G}\phi(x_1)\frac{\partial}{\partial x_2}\phi(x_2),\\
u_2&=&-\frac{\partial}{\partial x_1}\psi(x_1,x_2)=-\frac{1}{G}\phi(x_2)\frac{\partial}{\partial x_1}\phi(x_1).
\end{eqnarray}
\begin{figure}[t]
\centering
\includegraphics[width=0.7\textwidth]{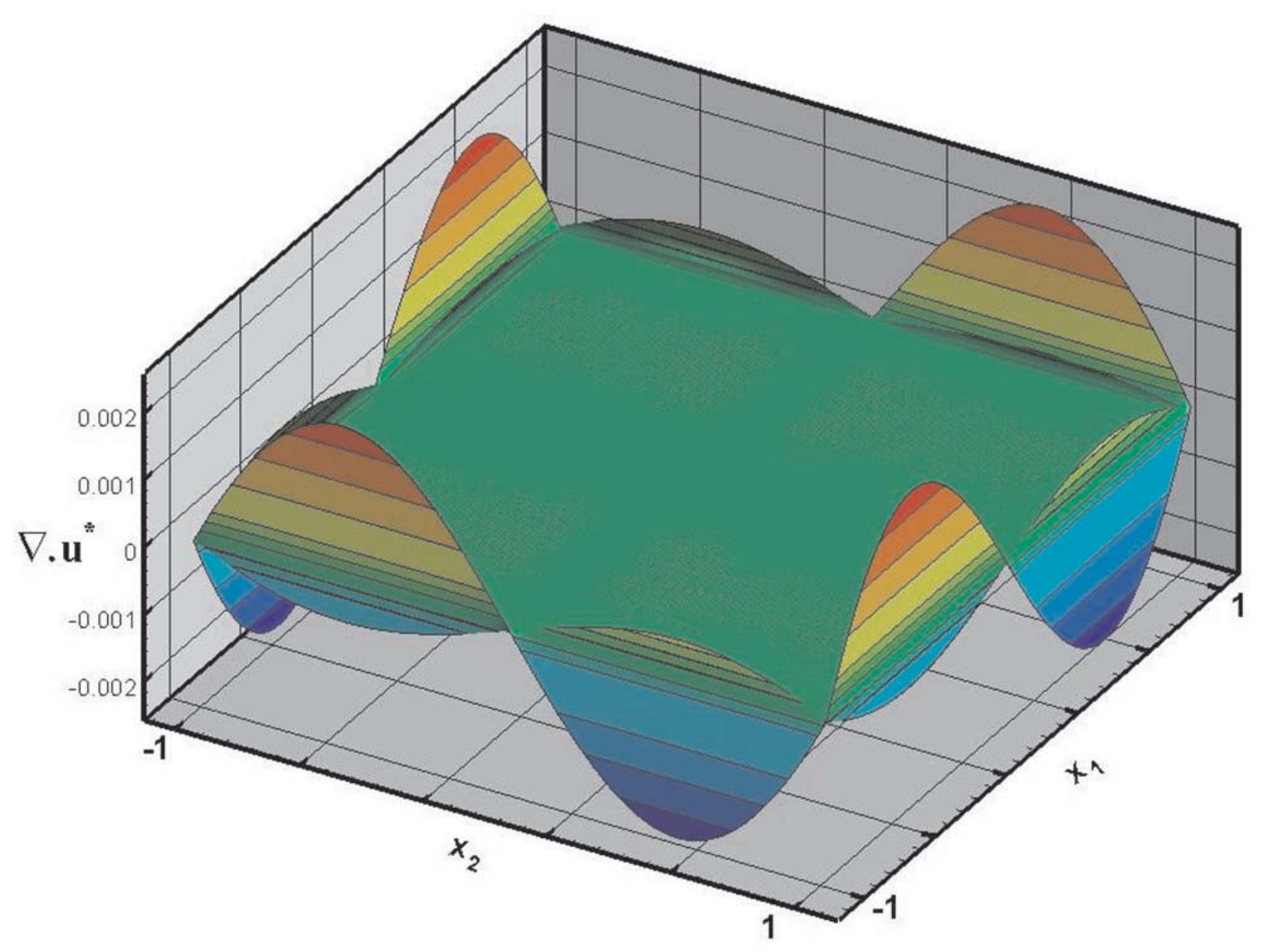}
\caption{Distribution of the divergence for the initial modified Chandrashkar--Reid flow. Although the velocities are obtained from a streamfunction, the divergence is not zero and its maximum values are occurred at the boundaries.}
\label{div_init}
\vspace{5mm}
\end{figure}
The left panel of Fig. \ref{vels_fig0} shows the above velocity filed on a $256^2$-point grid, which formally satisfies the no-slip conditions in both directions. In this configuration we defied $\bar{\Omega}_f=||{\bf x}||_{\infty}\leq 1$.\\
It is worth mentioning that although the above velocity vector is obtained from a streamfunction, it is not solenoidal. In fact, the divergence can be written as
\begin{equation}
\nabla\cdot{\rm{\bf u}}=\frac{1}{G}\left[ \frac{\partial}{\partial x}\left(\phi(x)\frac{\partial}{\partial y}\phi(y)\right)- \frac{\partial}{\partial y}\left(\phi(y)\frac{\partial}{\partial x}\phi(x)\right) \right],
\end{equation}
which obviously is not necessarily zero on the boundaries. For example, on the left boundary $\phi(x=-1)=0$, while $\frac{\partial}{\partial x}\phi(x=-1)\neq 0$.\\
This fundamental problem is one of the difficulties that the $\psi-\omega$ formulation is suffered from. In fact, in solution of $\nabla^2\psi=-\omega$, two sets of boundary conditions are needed to be satisfied (that is, $\psi(\Gamma_f)=\psi_{\Gamma}$, and $\frac{\partial}{\partial \hat{n}}\psi=0$),  which makes the problem over-determined \cite{Rempfer}. The divergence of the flow filed is shown in Figure. \ref{div_init}. As one can see, the divergences are not zero, especially right at the boundaries. \\
Now the method is applied to this flow. To this end, the solution domain is extended to a bigger regular domain  $D=\{||{\bf x}||_{\infty}<1.25\}$, and then the corresponding vorticity $\omega^*$ is found on $D$. The extended velocity field is shown in the right panel of Fig. \ref{vels_fig0}. Now Eqns. (\ref{Poissons}) are solved for the periodic as well as homogenous boundary conditions using the Fourier spectral, and the finite-difference methods; and the desired velocity ${\rm{\bf u}}$ is extracted from ${\rm{\bf u}}^*$. The distributions of $\nabla\cdot {\rm{\bf u}}$ for both cases are shown in Fig. \ref{vels_fig2}. As one can see, for the Fourier-spectral solutions the divergences are reduced to the machine accuracies (see the left panne of Fig. \ref{vels_fig2}). On the other hand, for the homogenous boundary conditions (which Eqns. (\ref{Poissons}) are solved using the  finite-difference method), although the divergences are no longer vanished completely, however, reduced appreciably. Table \ref{tab1} summarizes the maximum values of the divergences in the above cases.\\
\begin{figure}[t]
\centering
\includegraphics[width=0.48\textwidth]{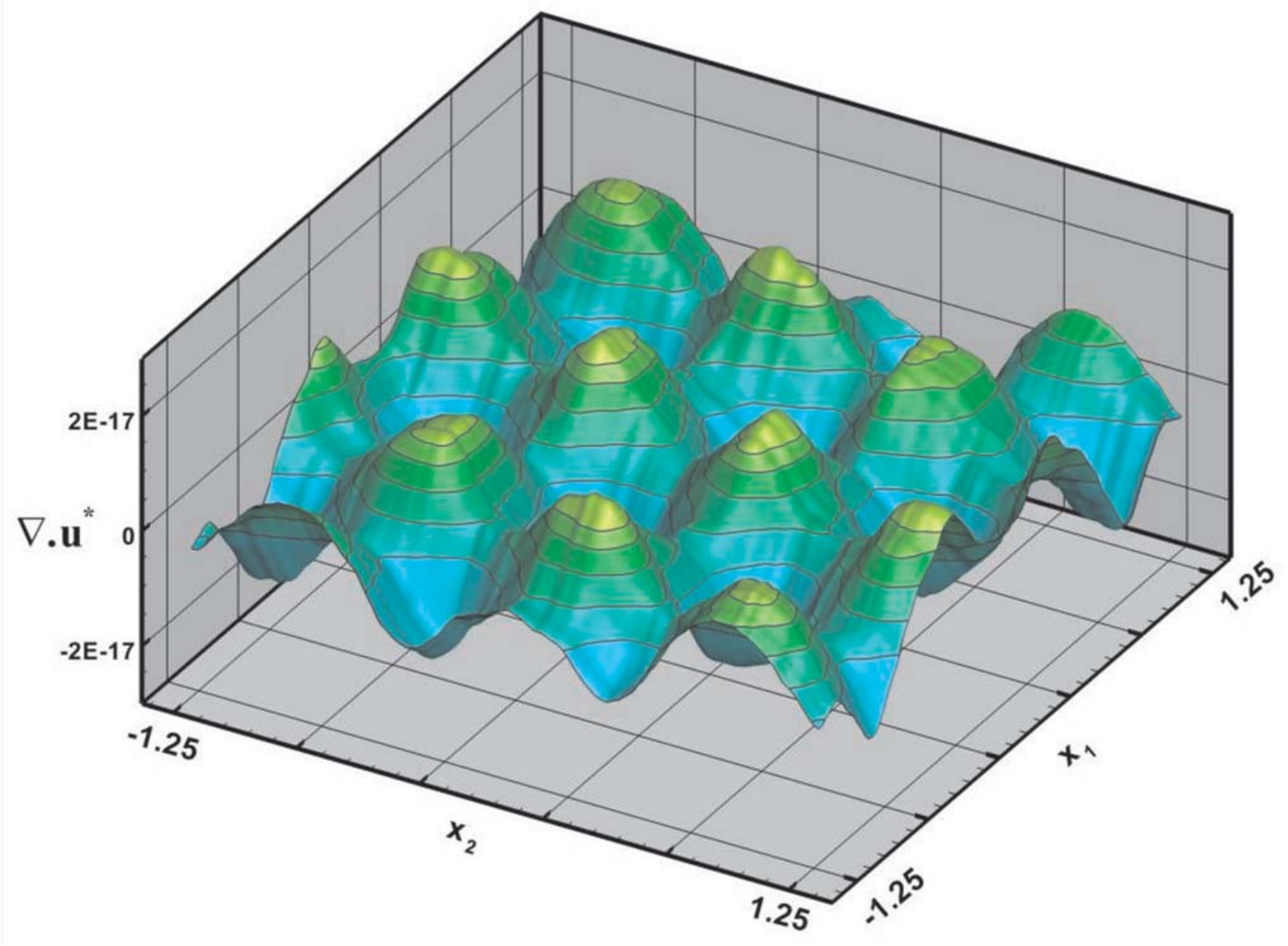}
\includegraphics[width=0.48\textwidth]{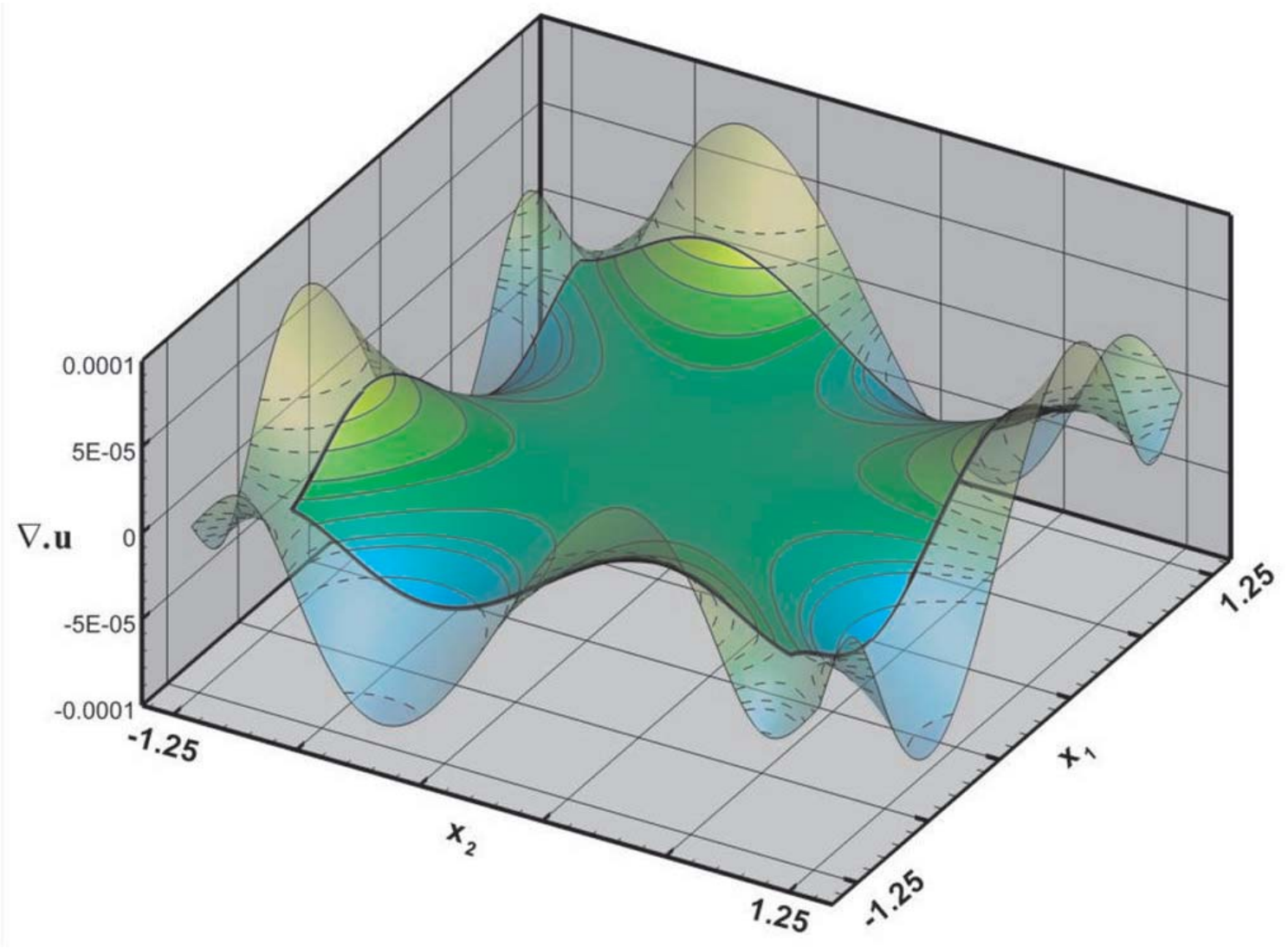}
\caption{The divergences for the extended modified Chandrashkar--Reid flow. {\bf Left:} Fourier spectral solution for the periodic boundary conditions on $\Gamma_D$. The divergence is reduced to the machine accuracy. {\bf Right:} Finite-difference solution  for the homogenous boundary conditions on $\Gamma_D$ (the dashed lines show the divergence of the extended velocity ${\rm{\bf u}}^D$, and  the solid lines show the desired final velocity ${\rm{\bf u}}$). The divergences are no longer within the machine accuracy, however, appreciable reduction is observable (c.f. Fig. \ref{div_init}).}
\label{vels_fig2}
\vspace{4mm}
\end{figure}
It is worth mentioning that although the above procedure improves the solenoidality of the flow, it may change the velocity boundary conditions. In fact, the velocities are changed all over the field, especially near the boundaries. This issue will be discussed in detail in the next section, because for the present test case boundary condition changes have not been sensible.
\subsection{Moving square cylinder}
In this section, the capability of the method in dealing with multiply-connected domains is examined. In this context, the practically important (and academically challenging) problem of a moving obstacle in a quiescent fluid is considered. Since the objective of the present paper is not implementation of the non-regular boundaries, a square cylinder, coinciding with the Cartesian grid, is considered. For the problem of non-regular immersed boundaries one can see the related references (see e.g. \cite{Sabetghadam1,Sabetghadam2}, or \cite{Balaras}). \\
\begin{figure}[h]
\centering
\includegraphics[width=0.9\textwidth]{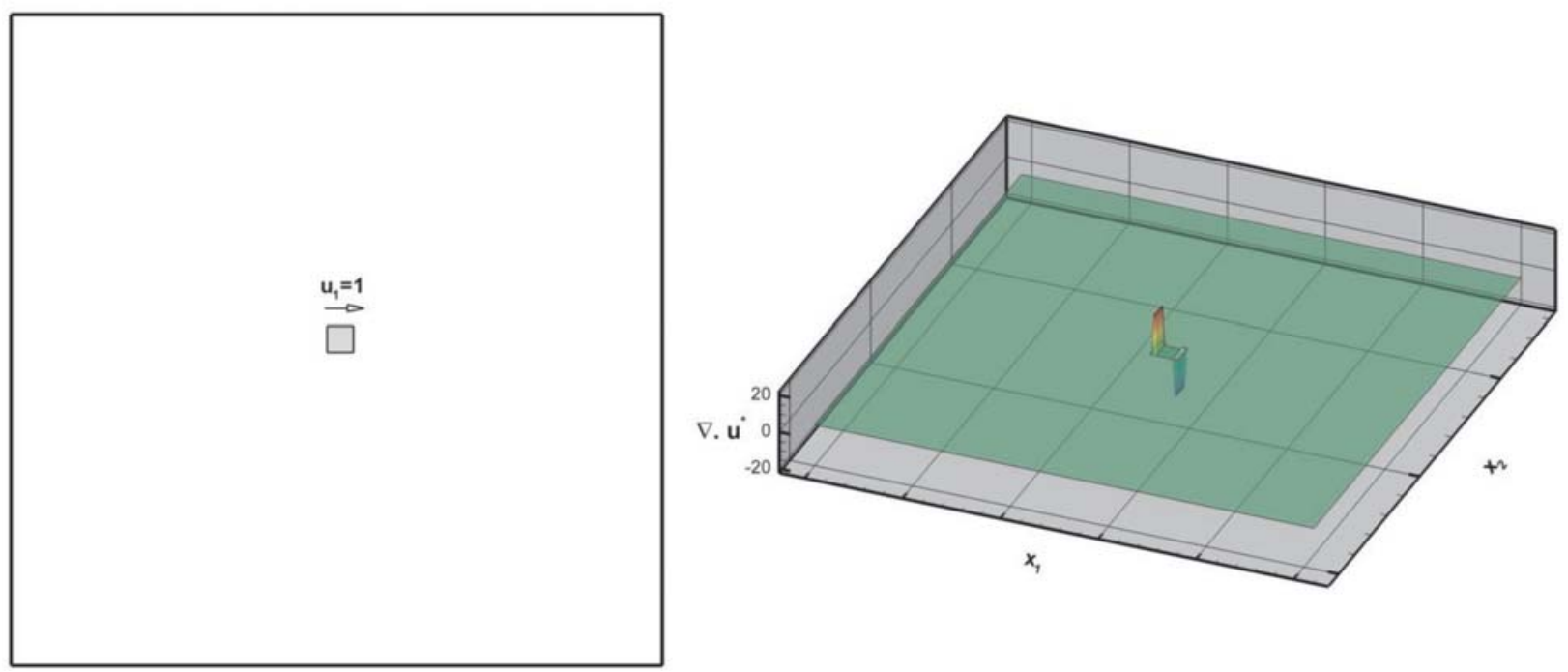}
\caption{ {\bf Left:} Moving square cylinder in a quiescent fluid. The immersed velocity boundary conditions $(u_1,u_2)=(0,1)$ are implemented sharply on the square cylinder $\Omega_s=\{ ||x_i-\pi||_{\infty} \leq 10/128\pi \}$, placed in the flow domain $[0,2\pi]$.  {\bf Right:} The resulting divergences. As one can see, the continuity is lost because of implementation of the boundary conditions.}
\label{fig6}
\vspace{6mm}
\end{figure}

\begin{figure}[t]
\centering
\includegraphics[width=0.9\textwidth]{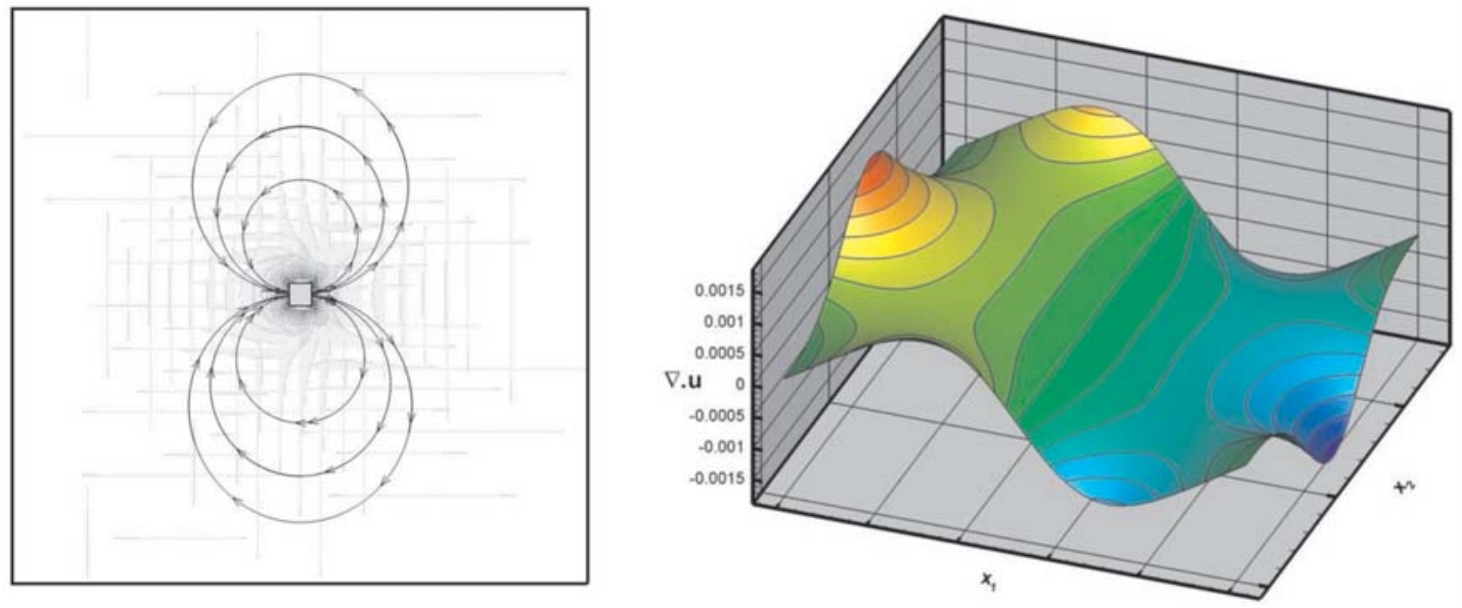}
\caption{The modified velocities for the moving square cylinder with homogenous boundary conditions. {\bf Left:}  The induced velocities are observable almost on the whole of the regular domain $D$. {\bf Right:} The maximum values of the divergences are reduced appreciably (c.f. Fig. \ref{fig6}), and occur on the outer boundaries.}
\label{fig7}
\end{figure}

Consider a regular fluid--solid domain $D$, consists of a square cylinder $\bar{\Omega}_s=\{ ||x_i||_{\infty}\leq 10/128\pi\}$, and the flow domain $\Omega_f=\bar{D}-\bar{\Omega}_s$. Moreover, assume that it is desired to implement the velocity boundary condition ${\rm{\bf u}}_s=(1,0)$ to the square cylinder (which will be treated here as an immersed boundary condition). Implementation of such a boundary condition imposes some discontinuities to the solution for $t\rightarrow 0$, the problem which is known as the corner singularity, or non-consistency of the boundary and initial conditions and were studied previously (e.g. see \cite{Temam1,Boyd, Temam2,Flyer}). The initial velocity and the distribution of the divergences are shown in Fig. \ref{fig6}, calculated on a $256^2$-- point grid. As one can see, the continuity is lost, especially on the boundaries of the moving obstacle, because of implementation of the sharp immersed velocity boundary conditions.\\
Now the method is applied to this flow for the homogeneous and periodic boundary conditions. It should be noted that since the velocities are already zero in the vicinity of the regular boundary $\Gamma_D$, the zero velocity margin $\cal B$ is not needed indeed. First we consider the homogeneous boundary conditions. In this case, the vorticity $\omega^D$ is found, and the Poisson's equations (\ref{Poissons}) are discretized on $D$ using the finite-difference method. The homogeneous Dirichlet  boundary conditions are implemented to the discretized equations, and the resulting linear problem is solved using the point successive over relaxation (PSOR) method. The modified velocity field and the divergences are shown in Fig. \ref{fig7}. As one can see in the left panel, the initial immersed non-solenoidal velocities are now spread across the domain $D$, such that the non-zero velocities are observable near the homogeneous outer boundaries (and therefore, the divergences are reduced near the square).
As a consequence, the divergences in the right panel show an appreciable reduction in the maximum values (with the factor of $\approx 10^{-4}$ in comparison to Fig. \ref{fig6}), and furthermore, these maximum values are now occurred on the outer (regular) boundaries.\\
\begin{figure}[t]
\centering
\includegraphics[width=0.9\textwidth]{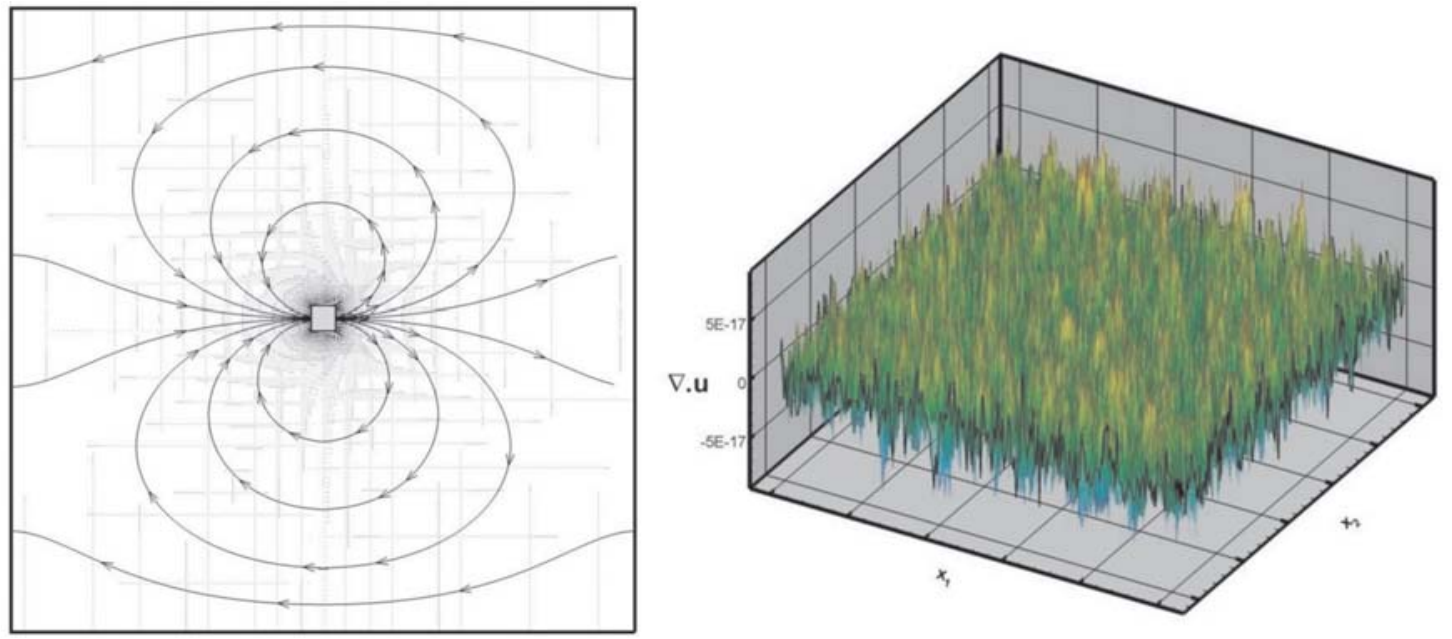}
\caption{The modified velocities for the moving square cylinder with periodic boundary conditions. {\bf Left:}  The periodic boundary conditions induced non-zero velocities on $\Gamma_D$ . {\bf Right}: However,  the velocity vector is solenoidal within the machine accuracies.}
\label{fig8}
\end{figure}
For the periodic boundary conditions, since the initial velocities ${\rm{\bf u}}^*$ are vanishing in the vicinity of the regular boundary $\Gamma_D$, the Fourier spectral solver may be used easily \cite{Boyd2, Sabetghadam1, Sabetghadam2}. The vorticity (on $D$) is obtained in the Fourier space, and the modified velocities are found via solution of Eqns. (\ref{Poissons}). The modified velocities and the divergences are shown in Fig. \ref{fig8}. As one can see, solenoidality of the velocity field is achieved up to the machine accuracy. With regard to the modified velocities (the left panel of Fig. \ref{fig8}), we should emphasized that since the periodic boundary conditions (not homogenous ones) are implemented, the modified velocities are not vanished on $\Gamma_D$ . In fact, non-zero velocities are induced on the $\Gamma_D$ by the moving obstacle. However, note that the overall mass balance is satisfied,  since the divergences are negligibly small.\\
\begin{figure}[h]
\centering
\includegraphics[width=0.9\textwidth]{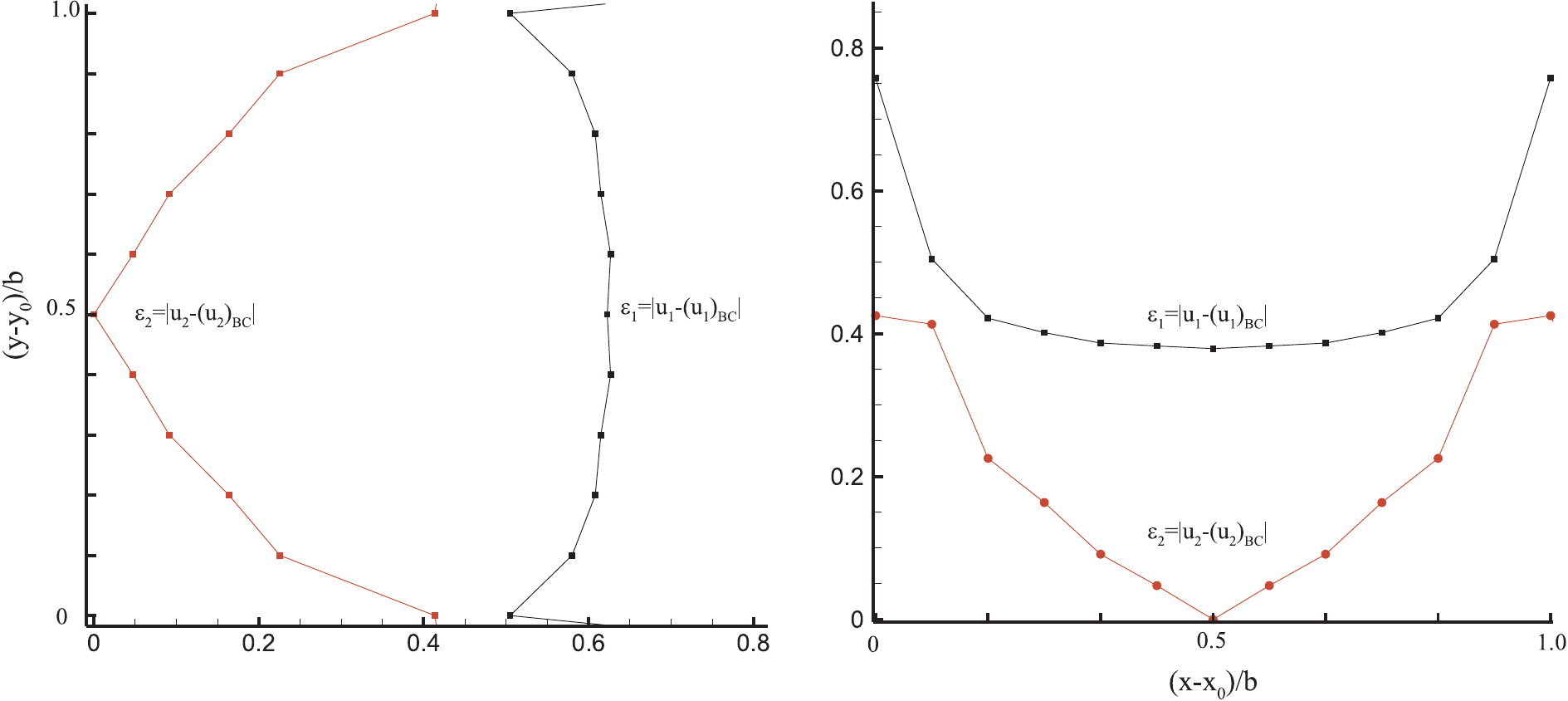}
\caption{Errors in immersed boundary conditions of the moving square cylinder due to imposing the solenoidality. The positions are presented with respect to the lower left corner of the square $(x_0,y_0)$.  {\bf Left:}  Errors in $u_1$ and $u_2$ on the left edge of the square cylinder. {\bf Right:} Errors on the upper edge of the square cylinder.}
\label{fig9}
\vspace{7mm}
\end{figure}
On the other hand, it should be emphasized that satisfying the solenoidality changes the immersed velocity boundary conditions.  Similar problem was observed in many other immersed boundary methods for the incompressible flow, and motivated emergence of some methods like the multi-direct forcing \cite{Wang1,Wang2}, or the implicit forcing method \cite{Le, Uhlmann} in the last years. To elucidate the issue, the modified velocities on the edge of the square cylinder are compared with the desired boundary conditions in Fig. \ref{fig9}. As one can see, both components of the velocity (i.e. $u_1$ and $u_2$) are changed on all the immersed boundaries. In fact this is the cost that we have to pay for working with physical (solenoidal) velocities. However, a body of evidence show that these discrepancies eliminates by developing the solution in the next times \cite{Sabetghadam2, Wang1, Wang2}; and furthermore, there are some remedies for overcoming this difficulty \cite{Sabetghadam2,Wang2, Uhlmann}.
\subsection{Extension to the three-dimensional problems}
Although merely the two-dimensional problems are considered up to now, the method may be extended easily to the three-dimensional flow. However, in these problems all three components of the vorticity vector present, and consequently we are faced with three Poisson's problems. Therefore, the method seems to be not so efficient (in comparison to e.g. some primitive variables formulations of the NSE). Nevertheless, since the boundary conditions, and the right hand sides of three Poisson's problems (\ref{Poissons}) are un-coupled, the problems are completely parallelizable. Therefore, the method has the potential of being competing, at least in principle. \\
\begin{figure}[t]
\centering
\includegraphics[width=0.9\textwidth]{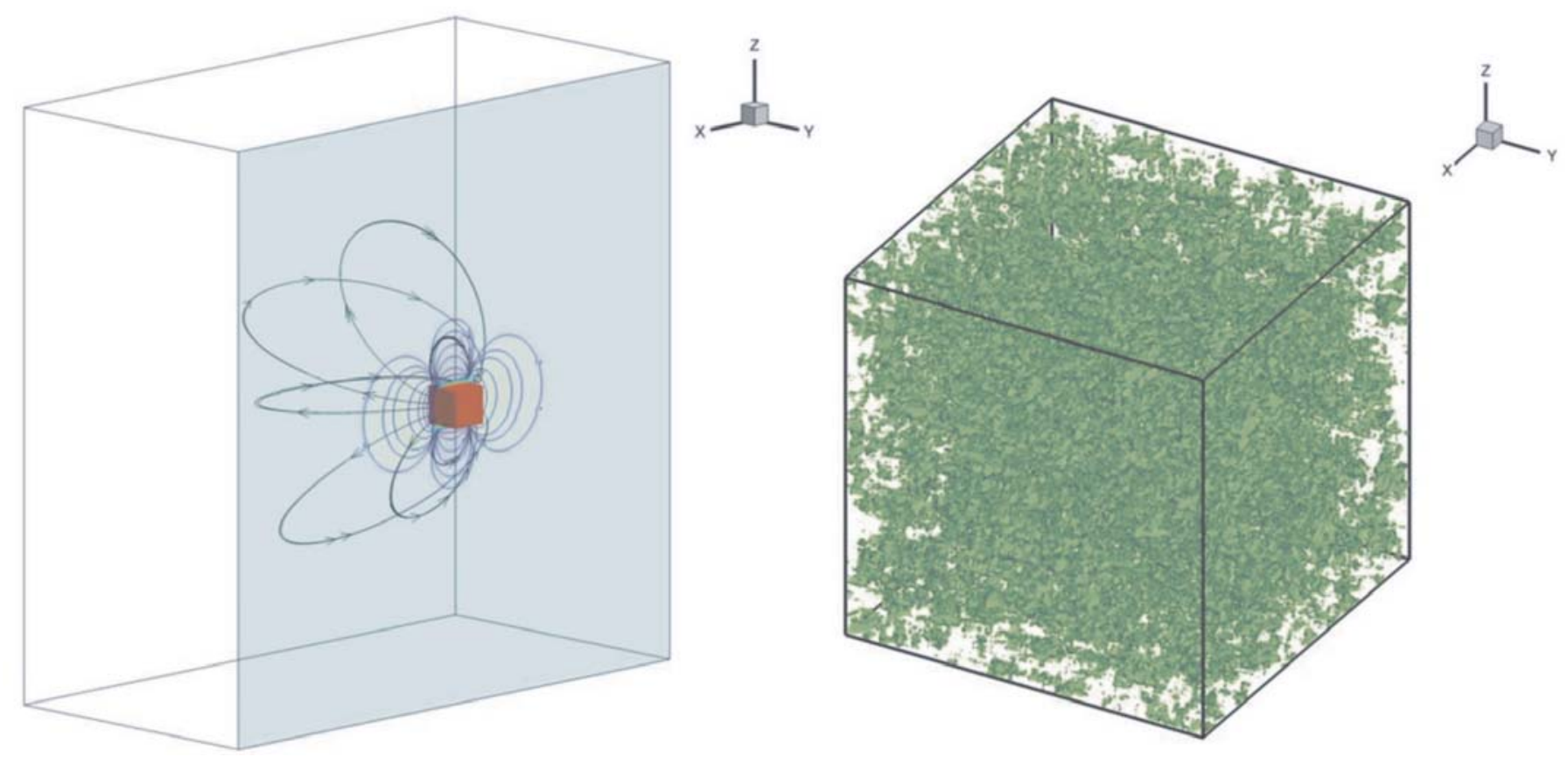}
\caption{The modified flow field for a moving cube (with velocity ${\rm{\bf u}}_s=(1,0,0)$) in the quiescent fluid with periodic boundary conditions.  {\bf Left:}  Streamlines and the contour plots of $u_1$ on half of the solution domain. {\bf Right:} Iso-surfaces of $10^{-17}$ level of the divergences of the flow. As one can see, the flow is divergence-free within the machine accuracy.}
\label{fig10}
\vspace{7mm}
\end{figure}
In the present work, our numerical experiment is restricted to the periodic boundary conditions (for which we can use the efficient Fourier--spectral solvers). The problem set up (which is an extension to the last moving square cylinder problem), is shown in Fig. \ref{fig10}.  A moving cube $\bar{\Omega}_s=\{ ||x_i-\pi||_{\infty}\leq 10/128\pi \}$ with velocity ${\rm{\bf u}}_s=(1,0,0)$ is placed in a regular solution domain $\bar{D}=\{||x_i-\pi||_{\infty}\leq 2\pi  \}$, occupied by a quiescent fluid. The extended vorticity vector is obtained and the modified velocities are re-calculated on a $128^3$ uniform grid, using the Fourier--spectral method. The results are shown in Fig. \ref{fig10}. As one can see, the cube velocity is induced to the whole of the solution domain, and the modified velocity vector is perfectly solenoidal.\\

\newpage
\subsection{Conclusions}
A method was suggested for construction of (nearly) solenoidal velocity vectors which satisfy (approximately) the desired immersed velocity boundary conditions. The method consists of re-calculation of the velocities from an extended vorticity field, obtained from an extended initial non-solenoidal velocity vector. It was shown that the solenoidality of the modified velocities are depended on the regular boundary conditions, and two different boundary conditions were considered. For the homogeneous boundary conditions, the method improves the solenoidality, and for the periodic boundary conditions (that is, the no-boundary problems), the solenoidality satisfies, up to the machine accuracy. The modified solenoidal velocities are no-longer satisfy the desired immersed velocity boundary conditions, however (as it is shown in other references), the discrepancies are eliminated by developing the solution. The method was applied to the two- and three-dimensional problems, and effects of the method on the velocity fields were discussed.
\newpage


%
\end{document}